\documentclass[preprint,review,12pt,authoryear]{elsarticle}

\usepackage{amssymb}
\usepackage{enumerate}
\usepackage{amsmath}
\usepackage{etoolbox}
\usepackage{url} 
\usepackage{hyperref}
\usepackage{enumitem}
\usepackage{upgreek}
\usepackage{environ}
\usepackage{algorithm}
\usepackage{algpseudocode}
\usepackage{setspace}
\usepackage{threeparttable}
\usepackage[margin=1in]{geometry}

\usepackage{lineno}

\journal {arXiv.org}

\begin{document}
\sloppy
\begin{frontmatter}



\title{Aggregating Google Trends: Multivariate Testing and Analysis}


\author{Stephen L. France\textsuperscript{a}}
\address{\textsuperscript{a}Corresponding Author: Mississippi State University, 40 Old Main, Mississippi State, MS, 39762, Email:sfrance@business.msstate.edu, Office: 662-325-1630}

\author{Yuying Shi\textsuperscript{b}}
\address{\textsuperscript{b}Texas A\&M Commerce, Commerce, TX, 75428, Email: Yuying.Shi@tamuc.edu}

\begin{abstract}
Web search data are a valuable source of business and economic information.  Previous studies have utilized Google Trends web search data for economic forecasting.  We expand this work by providing algorithms to combine and aggregate search volume data, so that the resulting data is both consistent over time and consistent between data series.  We give a brand equity example, where Google Trends is used to analyze shopping data for 100 top ranked brands and these data are used to nowcast economic variables.  We describe the importance of out of sample prediction and show how principal component analysis (PCA) can be used to improve the signal to noise ratio and prevent overfitting in nowcasting models. We give a finance example, where exploratory data analysis and classification is used to analyze the relationship between Google Trends searches and stock prices.  

\end{abstract}


\end{frontmatter}



\section{Introduction}
\label{}

Search data from the internet has rapidly become an important source of data for both analyzing trends in search terms and using these trends to predict underlying phenomena.  Google Trends has been widely used in the health care and epidemiology arenas \citep{REF1178}, particularly for using search terms to predict and monitor disease outbreaks \citep{REF1176}, for example, influenza \citep{REF1223}.  Other applications of Google Trends in these areas include examining the effect of tobacco control policies \citep{REF1208}, suicide incidence prediction \citep{REF1187}, and analyzing health screening behavior \citep{REF1228}.

In the economic literature, Google Trends data has been applied to ``nowcasting'', the process of forecasting current economic data before its release by using data available at a higher frequency than the economic data being predicted or available before the economic data are released \citep{REF1213,REF1214}.  Several researchers have investigated the use of web search volume data for economic prediction. \citet{REF1225} utilize keywords related to jobs and unemployment to predict the values in the weekly ``Initial Jobless Claims'' US government report.  \citet{REF1174} expand this work by also using Google Trends category searches to predict auto sales and Hong Kong tourism statistics.  Further nowcasting applications for Google Trends include forecasting labor and housing indicators in the United Kingdom \citep{REF1224}, auto sales \citep{REF1182,REF1183},  mutual fund flows from investor sentiment \citep{REF1218}, business cycles \citep{REF1209},  cinema admissions \citep{REF1179}, trading behavior \citep{REF1175},  private consumption \citep{REF1177}, issue salience in political science \citep{REF1180}, BitCoin prices \citep{REF1217}, and consumer sentiment \citep{REF1226}.  A certain commonality exists across most of these applications.  A set of words or topics is used to form a ``Google Trends Index'' and Google Trends values are calculated for the index across time.  The selected words or topics can be chosen either based on prior intuition or via an automated feature selection algorithm \citep{REF1181}. For the trends values to be useful, there is the assumption of a degree of  ``empirical similarity'' \citep{REF1227} between the trends index and the economic variable being predicted. A baseline auto-regressive AR(1) or AR(2) prediction model is created and then trend index terms are added to show prediction improvement from a baseline model. In such a prediction scenario,  \citet{REF1174} note the importance of using out of sample validation to ensure the correctness of the results. 

For most of above examples, the values of a search index term are measured relative to the same term over time. There is little analysis on how search terms for economic variables interact and change with respect to one another.  This is partially due to the limitations of the Google Trends platform.   Only a limited number of terms can be examined at any one time. Search values are relative only to values in the same search over time and different searches will have different absolute scales.  A major contribution of this paper is the development of a framework and associated algorithm for generalizing Google Trends results across multiple searches and creating multivariate data that have consistent values across series.  Several examples are given to illustrate the utility of this framework.  In the first example, Google Trends data for 100 major brands are used to predict a range of economic indices using nowcasting models.  In the second example,  Google Trends data are used to relate search volume to company financial performance and market structure using a combination of exploratory data analysis and time series clustering techniques.

\section{Aggregating Google Trends}
\label{}
There have been several attempts to create overall aggregate indexes from Google Trends data.  In comparing universities, \citet{REF1184} create a composite index of reputation by taking groups of five universities, ordering the universities based on Google Trends results, and then combining these partial orderings. However, this method does not result in a reliable quantitative index that is consistent over time.  In this paper, we utilize the concept of comparison items from \citet{REF1230} to develop algorithms to produce a quantitative Google Trends index.

\subsection{Algorithm}
In this section, we describe a set of algorithms for aggregating Google Trends searches.  Consider a trends search for two items over time periods $t=1\dots T$. Define the value of the trends index for the two items at time \textit{t} as $p_{1,t}$ and $p_{2,t}$ and let the underlying search volume be $v_{1,t}$ and $v_{2,t}$.  Search values are consistent inside a single Google Trends search, i.e., if there is some constant $k$ such that $v_{1,t}=k\left(p_{1,t}\right)$ then $v_{2,t}=k\left(p_{2,t}\right)$ and $r_{1,2,t}=\frac{v_{1,t}}{v_{2,t}}=\frac{p_{1,t}}{p_{2,t}}$, where $r_{1,2,t}$ is the ratio between views for items 1 and 2 at time \textit{t}. For a set of \textit{n} items, there are $\frac{n\left(n-1\right)}{2}$ possible item comparisons that can be estimated.

Google Trends reports only whole numbers on a scale of 1 to 100.  At a time period \textit{t}, the rounding error is bounded by $\frac{p_{1,t}-0.5}{p_{2,t}+0.5}\leq r_{1,2,t} \geq \frac{p_{1,t}+0.5}{p_{2,t}-0.5}$.  If the scores of both items are 50 then $0.9802 \leq r_{1,2,t} \geq 1.0202$, with an error of approximately $\pm 2\%$.  If, for example, one item completely dominates the other  and the item scores are 100 vs 1 then the error bound is $66.333 \leq r_{1,2,t} \geq 201$ giving a possible error of over 100\%.  Thus comparisons with similar items are preferable.  To mitigate the effects of error, the rationale behind our algorithms is to aggregate multiple similar comparisons for each item.  For items $i=1\dots n$ items, choose $j=1\dots m$ comparison items.  The comparison items can be internal or external to the \textit{n} items, but should cover the entire range of the \textit{n} items.  The number of items is a trade-off between the number of searches ($n \times m$) and the reduction in error from running multiple comparisons. 

Before running the algorithm, the Google Trends search should be run for each combination of item \textit{i} and comparison item \textit{j} across all time periods.  Further, define $p^{+}_{i,j,t}$ as the Google Trends value for item \textit{i} when compared with item \textit{j} at time \textit{t} and $p^{-}_{i,j,t}$ as the value of the comparison item \textit{j} when compared with item  \textit{i} at time \textit{t}.  The data are aggregated into two matrices ${\bf{S}}^{+} = (s^{+}_{i,j})_{\{n\times m \}}$ and ${\bf{S}}^{-} = (s^{-}_{i,j})_{\{n\times m \}}$, which hold the trends values summed across \textit{t}.  Here $s^{+}_{i,j}=\sum_{t=1}^{T}p^{+}_{i,j,t}$ and $s^{-}_{i,j}=\sum_{t=1}^{T}p^{-}_{i,j,t}$.

The Google Trends comparison algorithm is given as Algorithm 1. The algorithm sorts items with respect to volume from highest to lowest.  Ratios of views are then calculated between each item and the highest volume (base) item.  The algorithm then cycles through the items calculating views using the median comparison ratio between the current item and the base item. The median value is taken to prevent comparisons where one item dominates the other from influencing the results.  The base item is updated after $NC$ (number change) iterations. The choice of $NC$ trades off incremental error from the extremes of calculating the ratio of each item from the previous item and from scale/rounding error from comparing items with very different sizes.  

\begin{spacing}{2.0}
\begin{algorithm}
    \caption{Google Trends Combination Algorithm}
    \label{euclid}
    \begin{algorithmic}[1] 
        \Procedure{GoogleTrends}{$\bf{S}^{+}$,$\bf{S}^{-}$,$NC$} \Comment{\textit{NC} is the number of items to compare before reset.
        \textit{NC}}
            \State Ratio matrix ${\bf{R}}=\left(r_{i,j}\right)_{\{n\times m \}} \gets s^{+}_{i,j}/s^{-}_{i,j}\;\forall i,j$
             \State Sum plus vector ${\bf{SumPlus}}=\left(SumPlus_{i}\right){\{n\times 1 \}} \gets \sum_{j=1}^{m}s^{+}_{i,j}\;\forall i$
             \State Sum ratio vector ${\bf{SumRatio}}=\left(SumRatio_{i}\right){\{n\times 1 \}} \gets \sum_{j=1}^{m}s^{+}_{i,j}/\sum_{j=1}^{m}s^{-}_{i,j}\;\forall i$
            \State Aggregate ratings vector ${\bf{AgRatings}}=\left(AgRatings_{i}\right){\{n\times 1 \}} \gets 0$
             \State Multipliers vector ${\bf{Multipliers}}=\left(Multipliers_{i}\right){\{n\times 1 \}} \gets 0$
            \State Sort rows of $\bf{R}$ in descending order of ${\bf{SumRatio}}$.
            \State $i\gets 1,iBase\gets 1$  \Comment{Use Counter \textit{i} for row}
            \State ${\bf{AgRatings}}\lbrack 1\rbrack \gets 1000$
            \State  ${\bf{Multipliers}}\lbrack 1\rbrack \gets 1$
            \While{$i\leq n$} \Comment{Work through each row in turn}
            	  \State ${\bf{CompRatio}}\gets[]$ \Comment{Empty array of comparison ratios}
                  \For{$j=1:m$}  \Comment{Go through each comparison}
        			 \If{$r\lbrack i,j\rbrack/r\lbrack iBase,j\rbrack \in \mathbb{R}$}
        			     \State ${\bf{CompRatio}} \gets {\bf{CompRatio}}\cup \left(r\lbrack i,j\rbrack/r\lbrack iBase,j\rbrack\right)$ \Comment{If one of the values is not 0 or missing then add ratio}
        			 \EndIf
      		  \EndFor
      		  \State $ AgRatings\lbrack i\rbrack \gets AgRatings\lbrack iComp\rbrack \times median\left(CRatios\right)$ \Comment{Use the median ratio}
      		  \State $Multipliers\lbrack i\rbrack\gets\frac{AgRatings\lbrack i\rbrack\times{\bf{SumPlus}}_{iBase}}{AgRatings\lbrack iBase\rbrack\times{\bf{SumPlus}}_{i}}$ \Comment{Multipliers for row \textit{i} data}
      		  \State$i\gets1+1$  
      		   \If{$mod\left(i,NC\right)=1$} \Comment{Reset to the previous row}
        			  \State $iBase\gets i-1$     
        		  \EndIf
            \EndWhile\label{GoogleTrendsendwhile}
            \State \textbf{return} ${\bf{AgRatings}},{\bf{Multipliers}}$\Comment{Return the ratings and multipliers vectors}
        \EndProcedure
    \end{algorithmic}
\end{algorithm}
\end{spacing}

Initial experiments found that sizes around $NC=30$ gave good results on test datasets. The algorithm returns two vectors.  The \textbf{AgRatings} vector gives an overall index of popularity.  The \textbf{Multipliers} vector contains the multipliers to apply to the original trends data (on the selected comparison item), so that multivariate time series can be constructed across individual time points. This is important, as it allows consistent trends indexes to be built with any time period granularity the same or greater than the original data.

\section{Example 1: Developing a Brand Equity Index and Nowcasting}
The rationale behind the first example is to show how Google Trends data can be used to create a business based index that is economically useful.  In this example, Google Trends data were utilized to create several web search based brand equity indices for 100 leading brands in the US market.  These Google Trends based indices were then used to ``nowcast'' a range of consumer based economic indices.

\subsection{Brand Equity}
Consumer brands play an important role in both defining competitive markets and in driving the overall economy. In the marketing literature, there have been several conceptualizations of the idea of brand equity.  Brand equity has been defined at the microeconomic level as the utility that consumers have for a brand when the effects of marketing mix elements such as price and promotion have been discounted \citep{REF1464}.  In fact, from an individual consumer perspective, the effect of a marketing mix component will be more positive for a favorable brand than for an unfavorable brand \citep{REF1467}.

Brand equity can be defined from a financial perspective as the level of cashflow gained by branded goods over unbranded goods, which can be elicited from financial market valuations \citep{REF1465}.  Most practical measures of brand equity are multidimensional.  For example, \citet{REF1466} describes a measure that encompasses the areas of ``loyalty, perceived quality, associations, awareness, and market behavior'' and includes both financial and perceptual factors.  One widely used commercial measure, which we take as a starting point for our analysis, is the Interbrand (previously Interbrand-BusinessWeek) brand index, which has been utilized in past empirical work relating relative R{\&}D, advertising, and promotion spending to brand equity \citep{REF1468}. The 2017 edition of this index \citep{REF1469} utilizes  metrics derived from a range of data, including internal customer survey data, customer preference data, brand tracking data, and customer engagement data.  However, the overall valuation metrics are not revealed.  We took the list of the top 100 global brands from the list and used this to create several Google Trends web search indices.

\subsection{Exploratory Analysis}
\begin{figure}[!thb]
\centering
\includegraphics[scale=.52]{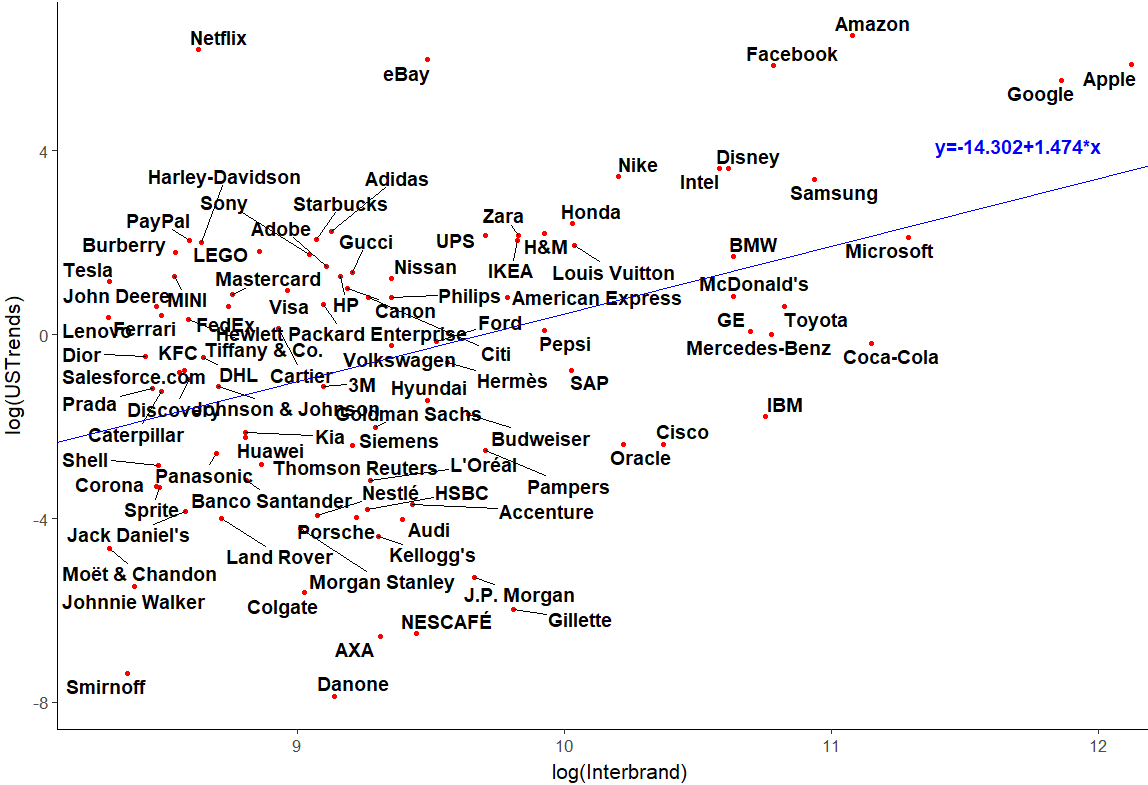}
\caption{Interbrands Value vs. US Overall Google Trends 2017}
\label{fig:PicAllRegression}  
\end{figure}

\begin{figure}[!thb]
\centering
\includegraphics[scale=.52]{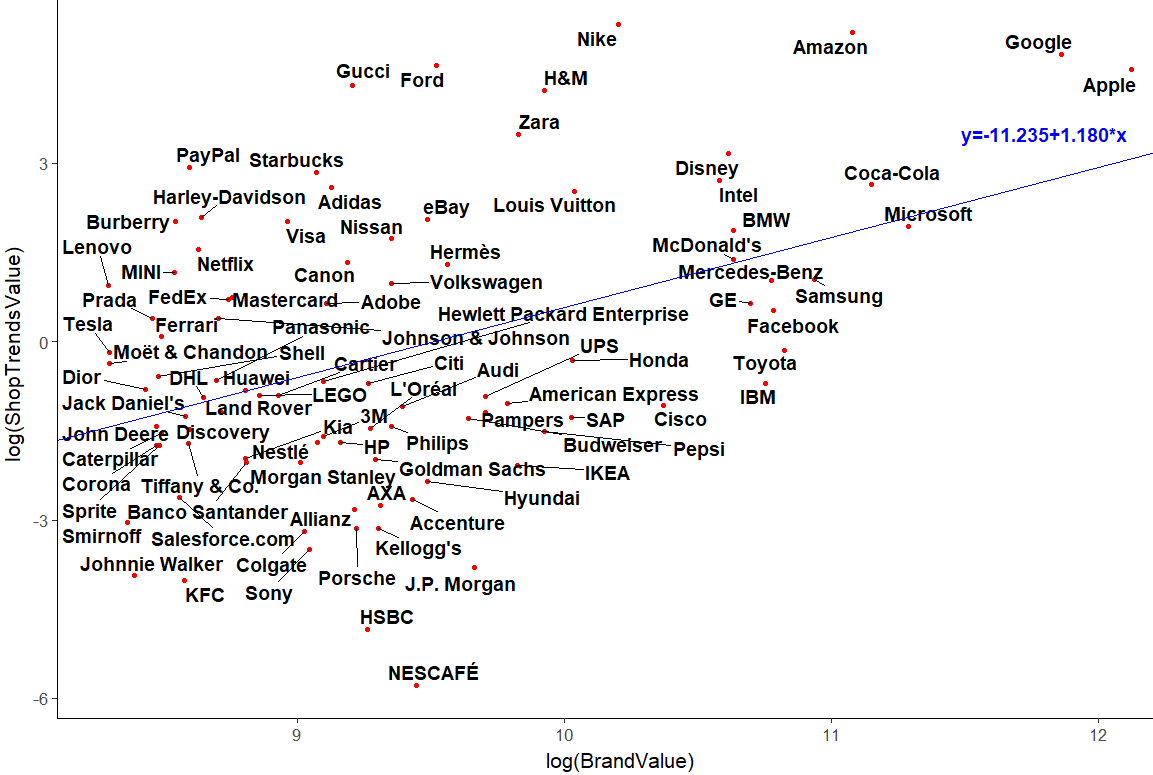}
\caption{{Interbrands Value vs. US Shopping Google Trends 2017}}
\label{fig:PicShopRegression}  
\end{figure}

Google Trends data for a ten year period  from Jan. 2008 to Dec. 2017 were taken for the 100 brands on the Interbrand list. The appropriate Google Trends code was selected for each brand and ten comparison brands were selected. Aggregate ratings were calculated using $NC=30$ and the multipliers were used to create a scale consistent multivariate time series. The combination algorithm was utilized to create a consistent multivariate index over time at the monthly level.  This process was repeated to create three different indices.  The first was a global index, for all countries.  The second was restricted to searches within the US.  The third was restricted to the US and to shopping queries, to give a ``purer'' measure of consumer sentiment.  Taking the aggregate trends data for 2017, the top twenty brand rankings for the Interbrand index and for the three trends indices are summarized in Table \ref{tb:BrandRankings}.
\begin{table}[!htb]
\centering
\caption{Brand Rankings Comparison}
\label{tb:BrandRankings}
\begin{tabular}{lllll} \hline
\textbf{Rank}&\textbf{Interbrand}&\textbf{World}&\textbf{US}&\textbf{US (Shop)}\\ \hline
1&Apple&Facebook&Amazon&Nike\\
2&Google&Google&Netflix&Amazon\\
3&Microsoft&Amazon&eBay&Google\\
4&Coca-Cola&eBay&Apple&Ford\\
5&Amazon&Samsung&Facebook&Apple\\
6&Samsung&Netflix&Google&Gucci\\
7&Toyota&Apple&Disney&H\&M\\
8&Facebook&IKEA&Intel&Zara\\
9&Mercedes-Benz&Adidas&Nike&Disney\\
10&IBM&Nike&Samsung&PayPal\\
11&GE&McDonald's&Honda&Starbucks\\
12&McDonald's&H\&M&Adidas&Intel\\
13&BMW&Nissan&H\&M&Coca-Cola\\
14&Disney&Honda&UPS&Adidas\\
15&Intel&BMW&Zara&Louis Vuitton\\
16&Cisco&Microsoft&Microsoft&Harley-Davidson\\
17&Oracle&Zara&Starbucks&eBay\\
18&Nike&PayPal&IKEA&Burberry\\
19&Louis Vuitton&Sony&PayPal&Visa\\
20&Honda&LEGO&Harley-Davidson&Microsoft\\ \hline
\end{tabular}
\end{table}
The results show strong face validity.  The Interbrand list and the two general Google Trends lists are dominated by the large global internet companies, such as Facebook, Amazon, and Google.  Apple is the top ranking brand in the Interbrand list, but is a little lower on the trends lists. The shopping only US list contains an array of consumer brands, including several that do not occur on other lists.  Nike is top of this list and does not appear higher than ninth in any of the other lists.  The correlations between the indices are given in Table \ref{tb:TrendsCorr}.  All correlations are significant, except for the correlation between the global trends overall index with the US trends shopping index, which has a smallish effect size ($r=0.1870$) and marginally significance ($p=0.0624$).  The Interbrand index is interesting, as its correlation with the US trends indices is slightly stronger than that with the global index, despite it being a global brand index.  The intercorrelations between the US overall, US shopping, and Interbrand indices are all in the range of 0.45-0.5, indicating that the Interband index may be intermediate to the overall trends index and the shopping index.
\begin{table}[!thb]
\centering
\caption{Interbrand and Google Trends Indices Correlations in 2017}
\label{tb:TrendsCorr}
\begin{tabular}{lllll} \hline
&\textbf{Interbrand}&\textbf{World}&\textbf{US}&\textbf{US (Shop)}\\ \hline
\textbf{Interbrand}&&$0.3462^{***}$&$0.4644^{***}$&$0.4773^{***}$\\
\textbf{World}&&&$0.4841^{***}$&$0.1870^{.}$\\
\textbf{US}&&&&$0.4910^{***}$\\ \hline
\end{tabular}
\begin{tablenotes}
\footnotesize
\item Using a t-test based on a Fisher transform, $^{.}$ $p<0.10$, $^{*}$ $p<0.05$, $^{**}$ $p<0.01$, $^{***}$ $p<0.001$.
\end{tablenotes}
\end{table}
To further explore the data and to gain insight into the US data, the trends values are plotted against the Interbrand values for the US overall search (Figure \ref{fig:PicAllRegression}) and the US shopping search (Figure \ref{fig:PicShopRegression}).  Least square regression lines are plotted for each figure. Natural log transforms are used on both variables to ensure error homoskedasticity and to prevent the regressions from being overly leveraged on the data from the large technology companies.
 One can see that for the US overall trends graph, the large technology companies have large positive residuals and predominantly business to business companies, such as Salesforce.com, Allianz, Accenture, and IBM, financial services companies, such as AXA, J. P. Morgan, and Morgan Stanley, and companies without significant online sales, such as KFC, have large negative residuals.  The pattern is similar for the US shopping trends graph, but with iconic consumer brands such as Nike and Ford, and apparel companies, such as H\&M, Zara, and Gucci having large positive residuals.  This perhaps indicates that US shopping trends series provides a stronger indicator of consumer strength and consumer buying power than the overall series.

\subsection{Nowcasting Consumer Economic Indices}
To help understand further how Google Trends brand information can be used for economic forecasting, we utilized the previously described trends series for the nowcasting prediction of several consumer based economic indicators.  These indicators, along with web links from the Federal Reserve Bank of St. Louis, are summarized  in Table \ref{tb:FedIndicators}.  This table gives the name of the series, the St. Louis Fed series code, the day range in the following month in which the data are typically released, and a description of the series.  All the series are monthly.  The series are either seasonally adjusted (SA in name) or unadjusted (UA in name). The series were chosen as having potential to be influenced by consumer brand searches and as having a release data after the end of the month covered by the data. The Michigan Consumer Sentiment series \citep{REF1475} is relevant, but is typically released within the covered month, so nowcasting on monthly Google Trends data would not help predict series values before release.  However, despite this fact, we include the series as a control, as it is probably the most widely used consumer sentiment index.
\begin{table}[htb]
\small
\centering
\caption{Summary of Nowcasting Economic Indicators}
\label{tb:FedIndicators}
\begin{tabular}{lllp{6cm}} \hline
Name&St. Louis Fed&Days&Description\\ \hline
ConMichUA&UMCSENT&0&University of Michigan index of consumer sentiment \citep{REF1501}.\\
ConOECDSA&CSCICP03USM665S&8-15&OECD Composite indicator of consumer sentiment \citep{REF1470}.\\
ConPriceSA&CPIAUCSL&11-18&Consumer price index of all urban consumers \citep{REF1471}.\\
ConPriceUA&CPIAUCNS&11-18&As above \citep{REF1472}.\\
PCEDGCSA&PCEDGC96&29-31& Real personal consumption expenditures on durable goods \citep{REF1473}.\\
UnempSA&UNRATENSA&2-9&Civilian unemployment rate \citep{REF1474}.\\ \hline
\end{tabular}
\end{table}
Another measure of consumer sentiment, the OECD composite indicator of consumer sentiment, is also included.  Consumer sentiment is strongly related to household spending, but in most cases, research has examined how lagged consumer sentiment can be used to predict household consumption or spending \citep{REF1478,REF1479,REF1476,REF1480,REF1481,REF1477}.  However, previous nowcasting research has shown that increases in web searches on products can precede sales \citep{REF1482}, so depending on the relative lags, changes in web searches for products could occur in the same period as changes in consumer sentiment.  In addition there may be a positive feedback effect \citep{REF1484}, where increases in consumer spending result in further increases in consumer confidence.  

Both seasonally adjusted and non-seasonally adjusted consumer price indices are included.  Consumer price inflation has a strong link to the level of demand from consumer purchases.  Classical economics, usually posits that inflation occurs to help bring a gap between demand and supply back into equilibrium and that inflation is affected by a range of factors including money supply, unemployment, production, and demand \citep{REF1486}. This relationship can hold in reverse. Inflation, along with interest rates and unemployment can affect the demand function for both durable and non-durable goods \citep{REF1487} and overall aggregate consumption \citep{REF1488}. Empirical work has generally found a small negative effect of inflation on attitudes towards spending  \citep{REF1489} and money allocated towards spending \citep{REF1490}. The psychological explanation behind this is that inflation causes consumers to worry about future purchases of essentials and thus cut back on current discretionary spending \citep{REF1492}.  At a more macroeconomic level, inflation is liable to trigger raises in interest rates to restrict the money supply, which consequently puts downward pressure on spending \citep{REF1493}. In summary,  with a bi-directional relationship between consumer purchase intention and inflation, there is a strong possibility that consumers' shopping searches could predict inflation in a nowcasting context.

Real personal consumption expenditures on durable goods is included and is a direct measure of consumer spending and could be predicted by shopping searches.  In addition, the seasonally adjusted employment rate is included.  As referenced in the previous paragraph, unemployment can affect demand for both durable and non-durable goods.  In fact unemployment can be thought of as a measure of ``aggregate income uncertainty'' \citep{REF1495} and along with consumer inflation can act as a measure of ``consumer discomfort'' \citep{REF1496}. Given previously described psychological considerations \citep{REF1492}, these factors are liable to affect consumer spending. The converse relationship can be true, with changes in consumer demand affecting unemployment and the relationship between unemployment and inflation \citep{REF1494}. Again, given a bi-directional relationship, there is a possibility that consumer searches could predict unemployment in a nowcasting context.

\subsection{Nowcasting Experimentation}
To test the use of the Google Trends brand data for nowcasting prediction, we follow the modeling approach of \citet{REF1225,REF1174}, but make methodological adjustments to allow for the fact that given the large number of data series for the 100 companies, any nowcasting model would be liable to overfit the data.  We thus implement methods designed to reduce the number of predictors and remove noise.

Consider a simple autoregressive model \eqref{eq:Baseline}, where $y_t$, an economic series value at time \textit{t} is predicted using the value of the series at time $t-1$. 
\begin{equation}
\label{eq:Baseline}
y_t=\beta_0+\beta_1y_{t-1}+\epsilon_t
\end{equation}
If the data are believed to be seasonal and have not been adjusted for seasonality then a term can be added for the period of the seasonality. For seasonal data, modeled at a monthly granularity level, a term for $y_{t-12}$ should be added to give \eqref{eq:BaselineS}.
\begin{equation}
\label{eq:BaselineS}
y_t=\beta_0+\beta_1y_{t-1}+\beta_2y_{t-12}+\epsilon_t
\end{equation}
Given \textit{n} Google Trends series, where series \textit{i} has value $g_{i,t}$ at time \textit{t}, a nowcasting prediction formula based on  \eqref{eq:Baseline} is given in  \eqref{eq:Full}.
\begin{equation}
\label{eq:Full}
y_t=\beta_0+\beta_1y_{t-1}+\sum_{i=1}^{n}\beta_{i+1}g_{i,t}+\epsilon_t
\end{equation}
The additional $y_{t-12}$ term from $\eqref{eq:BaselineS}$ can be added if adjustment for seasonality is required. Here, the monthly Google Trends data can be gathered immediately at the end of the month and used to generate an estimate of an economic series value before its delayed release.  However, such a formulation is likely to lead to overfitting and poor out of sample prediction.  Thus, several methods were implemented to reduce overfitting.  The first method employed was stepwise regression \citep{REF1497}. Here, a basic model is fit and then variables are incrementally added to improve the solution objective until it cannot be improved anymore.  The most common solution objective, which we utilize, is the Aikike information criterion \citep{REF702}, which is defined as $2k-2ln(L^*(\mu,\sigma^2))$, where $k$ is the number of model parameters, and $L^*(\mu,\sigma^2)$ is the regression maximum likelihood value. The lower the AIC, the better the fit relative to the number of parameters.  An overparameterized solution is penalized by the $2k$ term.  Forward stepwise regression  was implemented using the ``stepwise'' procedure in R.

The second procedure employed was the lasso \citep{REF509}.  The lasso works by restricting the total absolute value of the $\beta$ coefficients (excluding the intercept) in a regression, so in  \eqref{eq:Full}, $\sum_{i=1}^{n+1}|\beta_{i}|\leq\lambda$. This works in a similar fashion to ridge regression, where $\sum_{i=1}^{n+1}\left(\beta_{i}\right)^2\leq\lambda$.  Given the nature of the absolute value function, the lasso is  more likely to set individual $\beta$ values to zero, making it more suitable for independent variable selection.  The lasso was implemented using the glmnet package in R.  The optimal value of $\lambda$ for the data was found using 10 fold cross validation \citep{REF1372}. 

The third feature employed method was PCA (principal component analysis), a method of data reduction that summarizes data in a set of mutually uncorrelated new features or dimensions that maximize explained variance and minimize noise.  PCA has proved useful in a range of data summary and noise reduction applications \citep{REF1498}.  For example, Netflix ran a large scale data analysis competition for improving its movie recommendation engine.  Most entries utilized a method called collaborative filtering, which uses correlation patterns in reviews to make predictions.  Competition entrants soon found that applying PCA to the noisy, sparse review data increased prediction performance \citep{REF1499}.  Consider a matrix of trends values  ${\bf{X}}= (x_{ti})_{\{T\times n\}}$, where \textit{T} is the number of time periods in the data and \textit{n} is the number of brands in the dataset.  Let ${\bf{B}}={\bf{X\cdot}}{\bf{X\cdot}}'$, where ${\bf{X\cdot}}$ is mean centered to make ${\bf{B}}$ a covariance matrix and in addition can be standardized to make ${\bf{B}}$ a corrlelation matrix.  An eigendecomposition is performed, giving ${\bf{B}}={\bf{Q\Lambda Q'}}$, where $\Lambda$ contains the diagonalized eigenvalues, which give the proportion of variance accounted for by the data. A derived lower dimensional solution for \textit{k} dimensions is ${\bf{Y}}={\bf{Q\Lambda}}_k^{1/2}$, where the eigenvalues greater than \textit{k} in ${\bf{\Lambda}}_k$ are set to be 0. The columns of \textbf{Y} can then be used as input features in the regression model. Initial results showed that PCA on the covariance matrix (i.e., accounting for relative brand size) gave better results than PCA on the correlation matrix, so this method was used.

Given ten years of trends data, each data series consisted of 120 data points.  For the experiment, training data periods of $P\in\{30,60,90\}$ were utilized to build the model.  For each combination of \textit{P}, model, and dataset, the following procedure was followed. Starting with $t=2$ for seasonally adjusted data and $t=13$ for non-seasonally adjusted data requiring a seasonality term and going through to $t=T-P$, forecasting models were built and the model error was calculated using the mean absolute percentage error (MAPE), given in \eqref{eq:MAPE}.  The one-ahead out of sample  forecast  $\hat{y}_{t+1}$ was then calculated for each model and the MAPE value was calculated with respect to $y_{t+1}$.  The results were then averaged across time periods. Initial experiments showed that differencing the data and using log transforms of the dependent variables did not improve performance, so these data transformations were not performed.
\begin{equation}
\label{eq:MAPE}
MAPE=\frac{100}{n}\sum_{i=1}^n\left|\frac{y_i-\hat{y}_i}{y_i}\right|
\end{equation}

\subsection{Results}
The in sample results are given in Table \ref{tb:ResInSample} and the out of sample results are given in Table \ref{tb:ResOutSample}.
\begin{table}[!htb]
\centering
\caption{Brands Nowcasting: In Sample Results}
\label{tb:ResInSample}
\begin{tabular}{|l|l|l|lllll|} \hline
&&\textbf{MAPE}&$\Delta$\textbf{MAPE}&&&&\\ \hline
\textbf{TW}&\textbf{Row Labels}&\textbf{Base}&\textbf{Full}&\textbf{Fwd}&\textbf{Lasso}&\textbf{PCA1}&\textbf{PCA2}\\ \hline
30&ConMichUA&1.1092&\bf{-1.0293}&\bf{-0.9515}&\bf{-0.1640}&\bf{-0.0657}&\bf{-0.0664}\\
&ConOECDSA&0.0407&\bf{-0.0375}&\bf{-0.0348}&\bf{-0.0101}&\bf{-0.0020}&\bf{-0.0024}\\
&ConPriceSA&0.0433&\bf{-0.0398}&\bf{-0.0361}&\bf{-0.0065}&0.0001&0.0002\\
&ConPriceUA&0.0589&\bf{-0.0557}&\bf{-0.0527}&\bf{-0.0202}&0.0002&\bf{-0.0015}\\
&PCEDGCSA&0.1900&\bf{-0.1771}&\bf{-0.1724}&\bf{-0.0402}&\bf{-0.0026}&\bf{-0.0028}\\
&UnempSA&1.0553&\bf{-0.9435}&\bf{-0.8709}&\bf{-0.1918}&\bf{-0.0868}&\bf{-0.1130}\\ \hline
60&ConMichUA&1.0568&\bf{-0.9708}&\bf{-0.8691}&\bf{-0.1100}&\bf{-0.0729}&\bf{-0.0778}\\
&ConOECDSA&0.0402&\bf{-0.0367}&\bf{-0.0336}&\bf{-0.0064}&\bf{-0.0021}&\bf{-0.0021}\\
&ConPriceSA&0.0401&\bf{-0.0359}&\bf{-0.0339}&\bf{-0.0049}&\bf{-0.0001}&\bf{-0.0002}\\
&ConPriceUA&0.0571&\bf{-0.0530}&\bf{-0.0511}&\bf{-0.0184}&\bf{-0.0004}&\bf{-0.0019}\\
&PCEDGCSA&0.1711&\bf{-0.1564}&\bf{-0.1518}&\bf{-0.0299}&\bf{-0.0033}&\bf{-0.0011}\\
&UnempSA&1.0520&\bf{-0.9093}&\bf{-0.7888}&\bf{-0.1910}&\bf{-0.1084}&\bf{-0.1144}\\ \hline
90&ConMichUA&0.9848&\bf{-0.8128}&\bf{-0.6157}&\bf{-0.1444}&\bf{-0.0695}&\bf{-0.0832}\\
&ConOECDSA&0.0380&\bf{-0.0310}&\bf{-0.0258}&\bf{-0.0065}&\bf{-0.0025}&\bf{-0.0025}\\
&ConPriceSA&0.0379&\bf{-0.0294}&\bf{-0.0254}&\bf{-0.0034}&\bf{-0.0001}&\bf{-0.0002}\\
&ConPriceUA&0.0552&\bf{-0.0469}&\bf{-0.0441}&\bf{-0.0172}&\bf{-0.0012}&\bf{-0.0013}\\
&PCEDGCSA&0.1609&\bf{-0.1314}&\bf{-0.1223}&\bf{-0.0224}&\bf{-0.0012}&0.0003\\
&UnempSA&1.0762&\bf{-0.7909}&\bf{-0.5688}&\bf{-0.1686}&\bf{-0.0772}&\bf{-0.0792}\\ \hline
\end{tabular}
\end{table}
The TW column gives the number of periods in the training window.  The base column gives the average MAPE for the base autoregressive model.  The remaining columns give for each model the change in MAPE from the base model, so that a value less than 0 (bolded) indicates better performance than the base model.  The in sample results show that all augmented models, apart from a few  cases for the PCA models, gave improved in sample results.  The strongest reductions are for the full regression model, followed by the forward stepwise regression model. In these cases the negative $\Delta$MAPE values are almost as large as the MAPE for the base models, indicating virtually no in-sample error. However, it is likely that these models are overfit, a fact that can validated by looking at the out of sample results. 

For the out of sample results, neither the full or stepwise forward regression models gave a negative $\Delta$MAPE for any of the experimental conditions. The remaining three methods had out-of sample prediction improvements. The results were very dependent on the time window.  For example, for the unemployment data, for training windows of length 30 and 60, the lasso and both the one and two component PCA models gave improved out of sample prediction.  However, for windows of length 90, no technique managed an improvement in out of sample prediction.  Conversely, the one and two component PCA models give improved out of sample predictions on the seasonally adjusted consumer price data for the windows of length 90, but not for length 30 or 60. This indicates a trade-off between having enough data to accurately estimate the model vs. the immediacy of the data and possible changes in the economic environment introducing inaccuracy on models estimated with older data.  Interestingly, while the PCA models for unemployment and the Michigan sentiment index had quite strong out of sample performance (around 10\% reduction in MAPE from the base model), the models for the consumer price index and consumer expenditures on durable goods had more marginal performance. In the case of the consumer price index, this may be because other factors affecting consumer prices, such as interest rates and commodity prices outweigh any marginal effects from consumer spending. The consumer spending on durable goods results are a little more surprising.  Delving deeper into the data from this series( \url{https://fred.stlouisfed.org/release/tables?rid=54&eid=3220&snid=3217}), for January 2018, out of 13,743 billion of total consumer expenditures, 1,504 billion or around 11\% are listed as durable goods, while 9,324 billion or 68\% are services, indicating that consumer shopping searches would be a better proxy of a measure containing services spending rather than purely durable goods spending.
\begin{table}[!htb]
\centering
\caption{Brands Nowcasting: Out of Sample Results}
\label{tb:ResOutSample}
\begin{tabular}{|l|l|l|lllll|} \hline
&&\textbf{MAPE}&$\Delta$\textbf{MAPE}&&&&\\ \hline
\textbf{TW}&\textbf{Row Labels}&\textbf{Base}&\textbf{All}&\textbf{Fwd}&\textbf{Lasso}&\textbf{PCA1}&\textbf{PCA2}\\ \hline
30&ConMichUA&0.9053&38.2687&1.8464&0.0869&\bf{-0.0983}&\bf{-0.0857}\\ \hline
&ConOECDSA&0.0363&2.3190&0.0534&0.0054&\bf{-0.0023}&\bf{-0.0014}\\
&ConPriceSA&0.0335&3.1230&0.0788&0.0088&0.0023&0.0028\\
&ConPriceUA&0.0486&3.1174&0.0466&\bf{-0.0013}&0.0015&0.0036\\
&PCEDGCSA&0.1423&5.7609&0.3030&0.0639&\bf{-0.0023}&0.0007\\
&UnempSA&1.1954&71.6924&1.3947&\bf{-0.0889}&\bf{-0.1325}&\bf{-0.1050}\\ \hline
60&ConMichUA&0.7539&9.2272&1.9606&0.0385&\bf{-0.0692}&\bf{-0.1021}\\
&ConOECDSA&0.0301&0.7252&0.0476&0.0020&\bf{-0.0033}&\bf{-0.0029}\\
&ConPriceSA&0.0290&0.7648&0.1082&0.0093&0.0000&0.0002\\
&ConPriceUA&0.0424&0.8524&0.0741&0.0027&\bf{-0.0037}&\bf{-0.0015}\\
&PCEDGCSA&0.1140&3.2910&0.3798&0.0646&0.0249&0.0271\\
&UnempSA&1.3629&23.0648&1.5757&\bf{-0.0683}&\bf{-0.1189}&\bf{-0.0942}\\ \hline
90&ConMichUA&0.5475&8.3084&1.4952&\bf{-0.0394}&0.0502&\bf{-0.0154}\\
&ConOECDSA&0.0224&0.5208&0.0394&\bf{-0.0028}&\bf{-0.0004}&0.0005\\
&ConPriceSA&0.0293&0.2088&0.0962&0.0145&\bf{-0.0007}&\bf{-0.0017}\\
&ConPriceUA&0.0364&0.3826&0.0641&0.0074&\bf{-0.0050}&\bf{-0.0020}\\
&PCEDGCSA&0.0992&3.9579&0.4239&0.0616&0.0338&0.0284\\
&UnempSA&1.2990&24.6020&1.3155&0.0997&0.1419&0.1841\\ \hline
\end{tabular}
\end{table}

\section{Example 2: Exploratory Analysis of Financial Data}
The second example is in the financial domain.  Exploratory data analysis is used to explore the relationship between Google Trends results and stock market performance.  A key purpose of the analysis is to show how multivariate data that are both consistent over time and across individual series can be analyzed.

\begin{figure}[htb!]
\centering
\includegraphics[scale=.6]{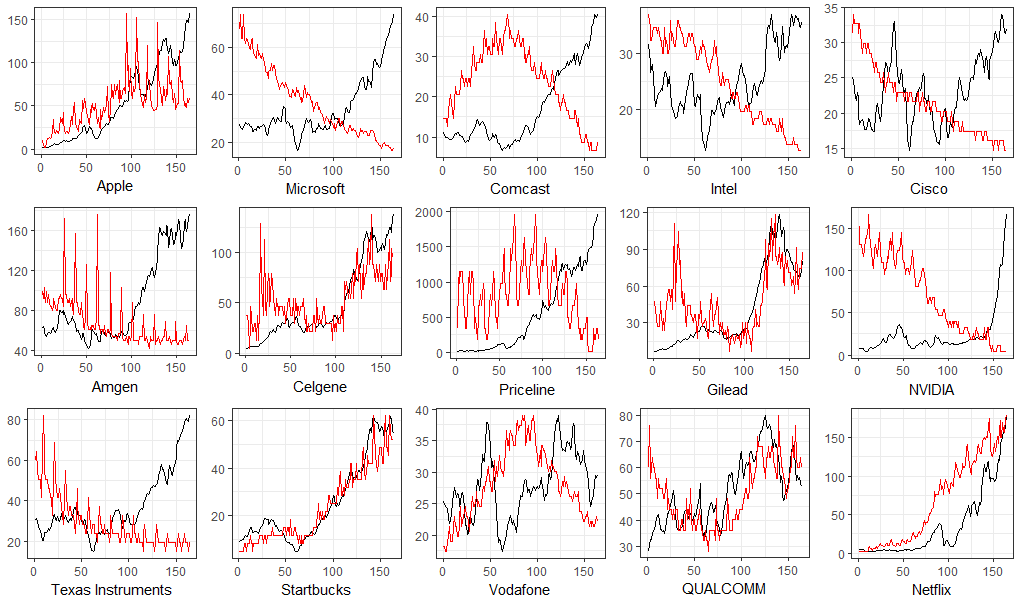}
\caption{Top 15 Companies (Black: Stock Prices. Red: Trends Values)}
\label{fig:Top15}  
\end{figure}

To start, monthly data were taken for the shares in the current NASDAQ 100 list.  The Google keyword for each company was used and eight comparison companies were chosen.  Aggregate ratings were calculated using $NC=30$ and the multipliers were used to create a scale consistent multivariate time series from Jan. 2004 to Aug 2017.  As a comparison, the average monthly stock prices were taken for each company in the same period (though data were not available for all of the shares for the entire period). Data are shown for the largest 15 companies with full data availability in Figure \ref{fig:Top15}.
One can see, for example, that Starbucks stock price grows with its trends popularity, but conversely, Microsoft's trends popularity decreases while its stock rises. To further examine patterns in the data, we examined the cross correlations between the trends values and stock price performance.  The results are given in Table \ref{tb:CrossCorr} . The first half of the table includes Pearson cross correlations for each company with lags (if positive, then the trends values lag the stock prices) of -5, -1, 0, 1, 5.  The mean value for each correlation is given, along with a count of how many companies have a correlation greater than 0 and a 2-tailed p-value, calculated from cumulative binomial distribution $X\sim bin(k,n=100,p=0.5),$ for how extreme the result is. In the second half of the table, partial correlations $r_{xy|z}$ are calculated, where $z$ gives the average monthly values for the NASDAQ 100 over the data period.  Here, accounting for overall stock market performance significantly increases the correlations between the trends series and the stock price series.

\begin{table}[htb]
\centering
\caption{Cross Correlations between Stock Market Prices and Trends Values}
\label{tb:CrossCorr}
\begin{tabular}{llllll} \hline
Corr Lag&-5&-1&0&1&5\\ \hline
Mean&-0.0113&-0.0406&-0.0534&	-0.0713&-0.0769\\
$k=\#(Corr>0)$&47&43&41&38&42\\
$2p\left(X\leq k\right)$&0.6173&0.1933&0.0886&0.0210&0.1332\\ \hline
Corr Lag&-5&-1&0&1&5\\ \hline
Mean&0.1221&0.1474&0.1373&0.1224&0.1258\\
$k=\#(Corr>0)$&69&69&69&67&67\\
$2p\left(X\geq k\right)$&0.0002&0.0002&0.0002&0.0009&0.0009\\ \hline
\end{tabular}
\end{table}

The overall calculated trends index is designed to have both relative consistency between items for a single time period and temporal consistency across time periods. Thus, multivariate time series methods relying on both absolute differences between series and differences between relative patterns can be employed.  To examine this, we performed time series clustering using the TSclust R package \citep{REF1232}.  In general, time series cluster analysis utilizes the same clustering algorithms as cluster analysis on cross-sectional data.  However distances are calculated using time series.  We utilized i) the Euclidean distance, which measures absolute differences between time series and ii) the Piccalo distance, which calculates the differences between the ARIMA models that best fit the series.  The Euclidean distance between time series \textit{x} and \textit{y} for time periods $1\dots T$, is given in \eqref{eq:Euclidean}.
\begin{equation}
\label{eq:Euclidean}
d\left(x_{1\dots T},y_{1\dots T}\right)=\left(\sum_{t=1}^T\left(x_t-y_t\right)^2\right)^{1/2}
\end{equation}
The Piccalo distance, given in \eqref{eq:Piccalo}, is defined using parameter estimates for AR($\infty$) processes, where the parameters for \textit{x} are defined as $\left(\hat{\pi}_{x1},\dots,\hat{\pi}_{xi},\dots,\hat{\pi}_{xk}\right)$ and the parameters for \textit{y} are defined as  $\left(\hat{\pi}_{y1},\dots,\hat{\pi}_{yi},\dots,\hat{\pi}_{yl}\right)$.  If $k<l$, then $\hat{\pi}_{xi}=0$ for $k<i\leq l$ and if $l<k$  then $\hat{\pi}_{yi}=0$ for $l<i\leq k$
\begin{equation}
\label{eq:Piccalo}
d\left(x_{1\dots T},y_{1\dots T}\right)=\left(\sum_{i=1}^{max\left(k,l\right)}\left(\hat{\pi}_{xi}-\hat{\pi}_{yi}\right)^2\right)^{1/_2}
\end{equation}

\begin{figure}[!h]
\centering
\includegraphics[scale=.6]{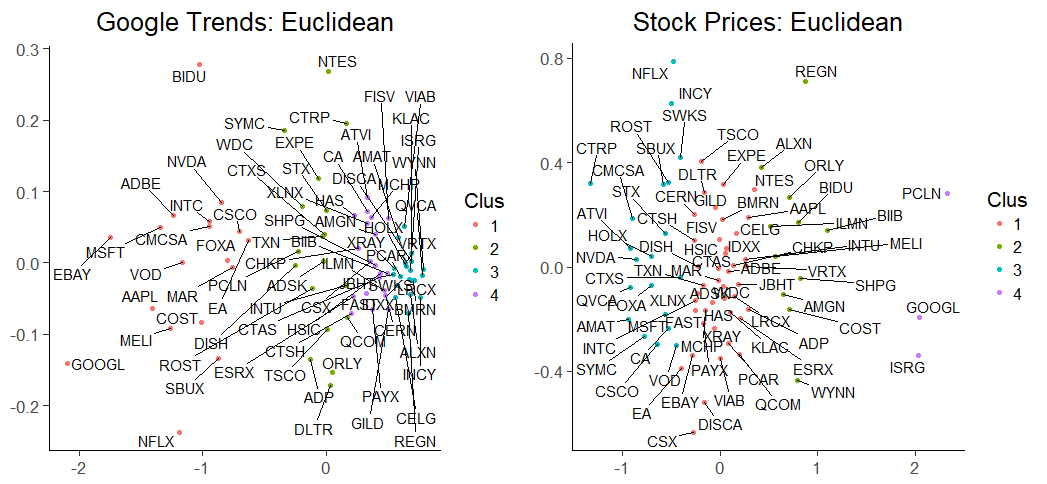}
\caption{Top 15 Companies (Euclidean Clusterings)}
\label{fig:EucTrends}  
\end{figure}

The analysis was carried out over the 10 years from September 2007 to August 2017. All companies ($n=77$) with at least 10 years of data were selected. The Euclidean distances were calculated on the natural log of the trends values to give a more compact scale representation.  The Piccalo distance requires stationary data. The augmented Dickey-Fuller test was run on the series and differenced series for both Google Trends and stock price data.  Results showed evidence that the series for the majority of companies were not stationary and that differenced series for all companies were stationary ($p<0.05$) for both trends and stock price series. Thus, differenced data was used as input for the Piccalo distance.  The resulting distance matrices were used as input to the k-medoids clustering procedure \citep{REF1366}, which is a form of partitioning clustering that is relatively robust to the problem of outliers.  This procedure was implemented using the ``pam'' function in R.  Four cluster solutions were chosen due to a relatively even silhouette plot and a high degree of interpretability. The results of the cluster analysis were plotted on a multidimensional scaling \citep{REF277,REF886} map, which was created from the same source distances using the ``smacof'' package in R \citep{REF452}.  The plots for the Euclidean distances are given in Figure \ref{fig:EucTrends} and the plots for the Piccolo distances are given in Figure \ref{fig:PicTrends}.
\begin{figure}[!h]
\centering
\includegraphics[scale=.6]{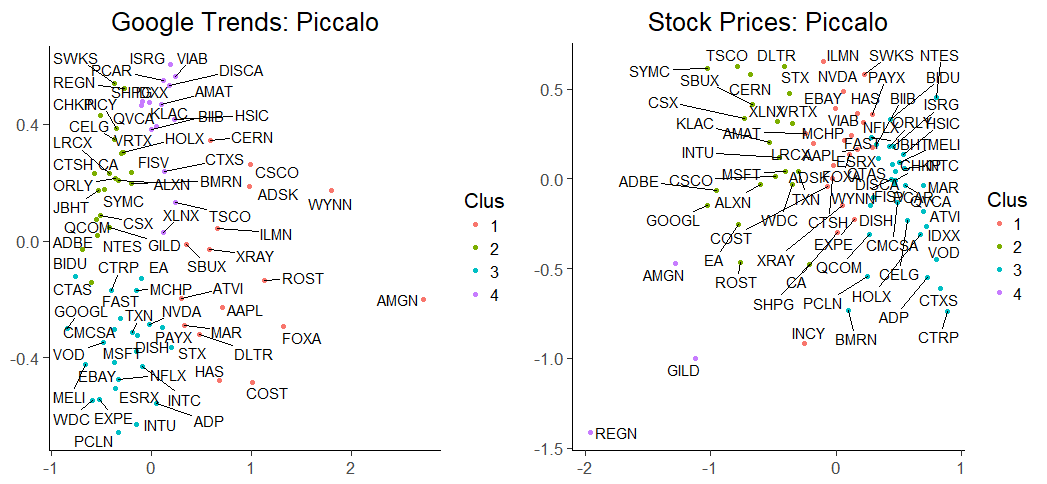}
\caption{Top 15 Companies (Piccolo Clusterings)}
\label{fig:PicTrends}  
\end{figure} 
The Euclidean (absolute) and Piccalo (temporal) representations produce significantly different mappings.  For example, in the Euclidean clusterings, Google is something of an outlier, due to high trends values and stock prices.  However, this is not the case in the Piccolo clusterings. In the Piccolo clusterings, companies are more liable to be clustered by category, particularly in the stock price clustering.  For example DISH (Dish Network), CMCSA (Comcast), and DISCA (Discovery Communications) are clustered very closely together due to being in the same sector and subsequently having similar stock price patterns.  To get a more macro view of the trends, time series graphs, given in Figure \ref{fig:PicTrends}, were created for each of the four clusterings. In each graph, the natural log values of the index or price for each item are plotted over the period of the analysis. The graph lines are colored by cluster.  One can see clearly that the ``absolute'' Euclidean clusterings are strongly homogeneous with respect to the index values, while the Piccolo clusterings are much more heterogeneous, which is to be expected given that these clusterings utilize temporal patterns.
\begin{figure}[!h]
\centering
\includegraphics[scale=.65]{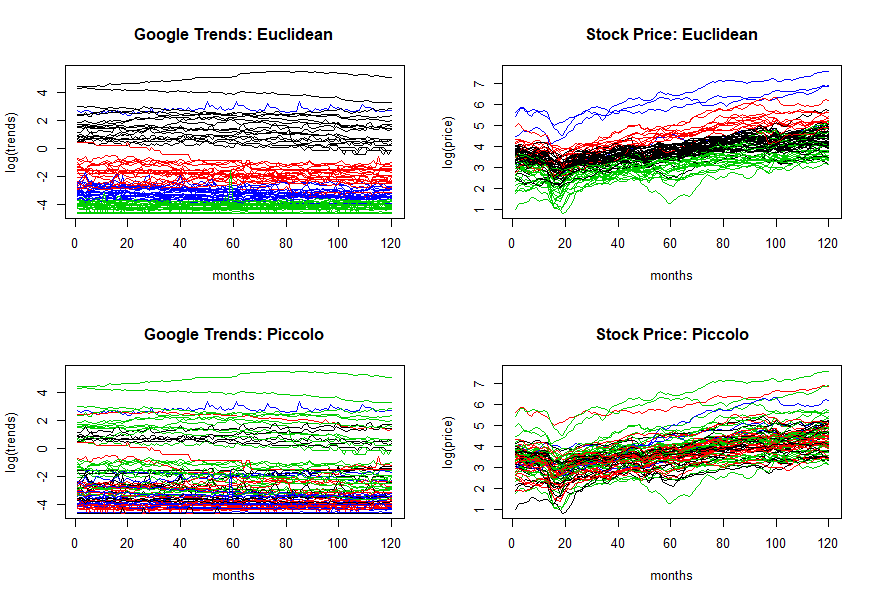}
\caption{Top 15 Companies (Time Series Clustering Summary}
\label{fig:PicTrends}  
\end{figure}

In conclusion, a range of exploratory analysis methods have been employed for this example.  Correlation analysis was used to show that while there is on average slightly negative correlation between Google Trends volume and stock market prices, when the overall direction of the market is taken into account, there is a significant positive correlation.  The multivariate analysis techniques of multidimensional scaling and cluster analysis were used to show the differences between market structure derived from relative changes in trends values and stock prices across time and market structure derived from absolute volume/price differences across time.

\section{Discussion}
\label{}
Google Trends web search data has been used in a wide range of economic and non-economic academic arenas.  We built on this and describe a heuristic algorithm for analyzing and combining Google Trends (or similar) search volume data across a large number of separate search queries.  We give a brand equity example, where we create several Google Trends based brand equity indices. We then utilize one of these indices for nowcasting a range of economic indices. We then give a financial example, where  multivariate Google Trends data are gathered for the companies in the NASDAQ index and show that with a trends index that is consistent over time, both absolute (i.e., differences between items) and temporal (differences over time) analyses can be performed.  Based on insight gained from the examples, a list of recommendations for the use of multivariate Google Trends data is given below.
\subsection{Google Trends Recommendations}
\begin{enumerate}
  \item Multivariate Google Trends data varies both across time and between the various levels of the different terms. The algorithm described in this paper can be used to ensure the consistency of the relative volume of multiple series over time.  This means that both methods that rely on consistency over time (e.g., clustering using correlations) and consistency between series (e.g., clustering using Euclidean distances) can be used.
  \item When performing nowcasting prediction, more specific searches give better results than general results, which can include queries made from many sources and with many different motivations.  When developing the nowcasting example, the shopping based search index gave better results than the general trends search.
  \item Getting improvement over the basic autoregressive model is  easy for in sample prediction, but difficult for out of sample prediction.  Data aggregated from a large number of search terms can contain noise and using these values as independent variables in a regression can lead to overfitting.
  \item Utilizing both the lasso and PCA on the covariance matrix gave positive results for out of sample prediction.  PCA performed better than the lasso. On the test data, one and two component PCA solutions gave improved predictions over the baseline model. Thus, PCA or a similar dimensionality reduction technique should be employed to de-noise data for out of sample prediction.
  \item The optimal time window length used for model building depends very much on the dependent variable being modeled.  There is a trade-off between having enough independent variable information and having only current information. For data analysis applications, the optimal time window length could be estimated empirically.
\end{enumerate}

\subsection{Future Work}
There are several avenues for future research.  The heuristic algorithm developed for this paper could be codifed and developed into a general purpose optimization algorithm.  Error bounds could be quantified and a general optimization formulation given.  Further work could be done in terms of visualizing and clustering Google Trends series and examining the properties of these series with respect to trend and random walk behavior.  Further nowcasting work could look to generalize the PCA methods utilized in this paper into a general partial least squares regression framework \citep{REF1500} that includes denoising with PCA and regression. 

The branding example given in this paper is designed to demonstrate the use of Google Trends based nowcasting in a specific domain.  It produced some insights into how brand searches are related to different economic variables, but a more in depth theory based study could examine the relationship between different types of brand searches, the brand category, and economic variables (macro) or company performance (micro).  The analysis of electronic consumer word of mouth (eWOM) using social media or online review websites has been a very popular avenue of research in both marketing and information systems over the past few years; see for example, \citet{REF1093,REF831,REF1215,REF679,REF804}. Combining eWOM data with Google Trends web search data could provide insights not available when purely using one of these sources of data.  In particular, nowcasting models that incorporate both search popularity from trends data and consumer sentiment from eWOM data could explain a wide range of consumer and economic behavior.

\section*{Acknowledgments}
\label{}
This research did not receive any specific grant from funding agencies in the public, commercial, or not-for-profit sectors.



\section{References}
\begin{spacing}{1.0}
\label{}
\bibliographystyle{model5-names}
\bibliography{ELTrendsCompetitiveAnalysis2}
\end{spacing}






\clearpage

\end{document}